\documentclass[12pt]{article}
\usepackage{amsmath,amssymb,exscale,multirow,graphicx}
\usepackage[utf8]{inputenc}
\usepackage{xcolor}
\usepackage{cite}
\usepackage{hyperref}
\usepackage{subcaption}
\usepackage{float}

\newcommand{\m}[1]{\marginpar{{\tiny *}} }
\newcommand{\Fslash}{{\not \!F}}
\newcommand{\Order}{\mathcal{O}}

\begin{document}
\topmargin -1.0cm
\oddsidemargin -0.8cm
\evensidemargin -0.8cm

\begin{center}
\vspace{40pt}
\Large \textbf{A composite Froggatt-Nielsen model of flavor}
\end{center}

\vspace{15pt}
\begin{center}
{\bf Leandro Da Rold$^{\circ}$, Federico Lamagna$^{\star}$} 

\vspace{20pt}

\textit{Centro At\'omico Bariloche, Instituto Balseiro and CONICET}
\\[0.2cm]
\textit{Av.\ Bustillo 9500, 8400, S.\ C.\ de Bariloche, Argentina}

\end{center}

\vspace{20pt}
\begin{center}
\textbf{Abstract}
\end{center}
\vspace{5pt} {\small \noindent
A natural composite Higgs demands the presence of light resonances at the TeV scale, that in general are in conflict with bounds from flavor and CP violation. We propose a composite model with a Froggatt-Nielsen mechanism, that offers new possibilities for the origin of flavor. We analyse the interplay of partial compositeness and the horizontal U(1) symmetry in achieving the quark masses and mixing angles. We study the contributions to $\Delta F=2$ 4-fermion operators, as well as to $\Delta F=1$ and neutron dipole operators. We find scenarios in which the contribution to Left-Right and Right-handed operators involving the first and second generations can be suppressed, in particular for a region of parameter space it is possible to simultaneously suppress the mixed-chirality contribution to $K^0-\bar K^0$ mixing by one power of the Cabibbo angle, $\lambda_C$, and the dipole moments by $\lambda_C^2$ compared with anarchic partial compositeness, possibly making the resonances accessible at LHC. 4-fermion operators of $B_s$-meson mixing and Left-handed operators are not suppressed.
}

\vfill
\noindent {\footnotesize E-mail:
$\circ$ daroldl@ib.edu.ar,
$\star$ federico.lamagna@cab.cnea.gov.ar
}

\noindent
\eject


\section{Introduction}
The discovery of the Higgs boson has definitely established the Standard Model (SM) as the best description nowadays of the elementary particles and interactions. The quarks of the SM show a particular pattern of masses and mixing angles, with ratios that at the TeV scale~\cite{Antusch:2013jca} can be parametrised in terms of the Cabibbo angle $\lambda_C$ as:
\begin{align}
& m_u:m_c:m_t \sim \lambda_C^8:\lambda_C^4:1 \ , \nonumber\\
& m_d:m_s:m_b \sim \lambda_C^5:\lambda_C^3:1 \ , \nonumber\\
& m_b:m_t \sim \lambda_C^3:1 \ , \nonumber\\
& (V_{\rm CKM})_{12} \sim \lambda_C \ , \quad
(V_{\rm CKM})_{13} \sim \lambda_C^3 \ , \quad
(V_{\rm CKM})_{23} \sim \lambda_C^2 \ .
\label{eq-mbovermt}
\end{align}
In the SM this pattern is generated at high energies by the Yukawa interactions. However, if the SM is an effective theory valid up to a scale much larger than the TeV, the Higgs potential suffers from the hierarchy problem of the electroweak (EW) scale. One of the most attractive solutions to this problem is to consider that the Higgs is a composite state of a new sector that is strongly coupled at the TeV scale. In this case the Yukawa couplings depend on the type of interactions between the SM fermions and the new sector, and the flavor structure must be generated at energies much smaller than the Planck scale~\cite{Panico-Pomarol}.

One of the most interesting paradigms for the generation of flavor in composite Higgs theories is partial compositeness, in which the SM fermions are elementary fields that interact linearly with the operators of the new strongly interacting sector, with bilinear interactions being highly suppressed~\cite{Kaplan:1991dc}. At low energies the linear interactions lead to mixing between the elementary fermions and the composite resonances, generating Yukawa couplings with the composite Higgs that are controlled by the mixing. If the theory at high energies does not have any flavor symmetry the partial compositeness is anarchic (APC), in which case the hierarchy of linear mixing gives a rationale for the hierarchy of masses and mixing angles.
Since flavor transitions are also controlled by this mixing, APC contains a built-in GIM mechanism. Still, constraints from the CP violating observable in the $K$ meson system, $\epsilon_K$, as well as mixing of $B$-mesons and neutron dipole moments, require the scale of compositeness to be roughly one order of magnitude above the TeV, reintroducing a small tension with the stability of the EW scale. The most stringent constraints arise from the Wilson coefficient of the Left-Right operator $(\bar d_Rs_L)(\bar d_Ls_R)$, that requires resonances with masses of 10-20 TeV~\cite{Csaki}, and from dipole operators of quarks of first generation, that require mass over coupling of resonances of order 5 TeV~\cite{Neubert}. Flavor and CP symmetries can alleviate these issues~\cite{Santiago:2008vq,Csaki:2008eh,Redi-Weiler}, but they are in tension with LHC tests of compositeness of the light quarks~\cite{1201.6510}. More promising is the proposal of dynamical flavor scales~\cite{Panico-Pomarol}, as well as the addition of tiny bilinear interactions~\cite{1708.08515}.

A very different approach to flavor is the mechanism of Froggatt-Nielsen (FN)~\cite{Froggatt:1978nt,Davidson:1979wr}. By the introduction of a U(1) flavor symmetry and the choice of suitable charges for the SM fields, FN generates Yukawa interactions from higher dimensional operators, with a well defined power counting, that can lead to the pattern of Eq.~(\ref{eq-mbovermt})~\cite{Bordone:2019uzc}. Several details of this scenario depend on the dynamics of the new sector, such as the mechanism that triggers spontaneous breaking of the U(1) symmetry, the scale of this breaking and the phenomenology of the associated axion. Usually FN requires many heavy fermions~\cite{Babu:2009fd}.

In this paper we propose to combine partial compositeness with FN, by imposing a U(1) flavor symmetry in the composite sector, spontaneously broken by the strong dynamics and respected by the mixing with the elementary fermions. The flavor structure of the light fermions, that are identified with the SM ones, depends on the size of the elementary-composite mixing, as well as on the U(1) charges, such that the model allows to interpolate between APC and FN. While the top quark mass requires a considerable degree of compositeness and dimension four Yukawa interactions, the masses of the light quarks can be suppressed by a small mixing, as well as by the Wilson coefficient of the higher dimensional Yukawa interaction, that is determined by the U(1) charges. The Left- and Right-handed mixing angles of the SM fermions also follow a well defined pattern determined by mixings and charges. This interplay opens up new possibilities, in particular for dimension six flavor violating operators that mediate $\Delta F=1$ and $\Delta F=2$ transitions, as well as for the flavor diagonal and CP violating dipole operators. We will show that for some scenarios it is possible to suppress the Wilson coefficient of 4-fermion operators involving light Right-handed quarks, as well as those of dipole operators. We will consider in this work only the sector of quarks. 

Related attempts in the lepton sector have being proposed in~\cite{Bazzocchi:2003ug}, whereas not fully realistic proposals in the quark sector were presented in Refs.~\cite{Bazzocchi:2003vh,Bazzocchi:2004qx,Bazzocchi:2004cv}. At the level of effective field theories some interesting attempts have been proposed within two Higgs doublet models in~\cite{Bauer:2015fxa,Bauer:2015kzy}. Ref.~\cite{Alanne:2018fns} has considered an interesting model with a U(1) horizontal symmetry in which the Higgs and the axion are pseudo Nambu-Goldstone composite states. A more related proposal was recently given in Ref.~\cite{Chung:2021fpc}, although a general analysis of flavor constraints is missing.

The paper is organised as follows, in sec.~\ref{sec-model} we describe the basic idea and introduce a model with a content of composite resonances of the strong sector that can generate the flavor structure of the quarks. We show a set of solutions that reproduce the quark masses and the CKM matrix, and study the interactions with the composite resonances that can induce flavor violating processes. In sec.~\ref{sec-constraints} we show the predictions for flavor and CP violating operators, comparing them with APC. The tables shown on this section contain the most important results of the paper. In sec.~\ref{sec-other} we discuss very briefly constraints from dijets and in sec.~\ref{sec-axion} we show some general features of the interactions of the axion of the theory. Finally, we raise some discussions and conclude in sec.~\ref{sec-conclusions}.

\section{A model of flavor with partial compositeness and \\Froggatt-Nielsen mechanism}\label{sec-model}
Our proposal is similar to anarchic partial compositeness (APC), where the SM fermions are elementary states that at a high energy scale $\Lambda_{\rm UV}$ have linear interactions with the operators of a new strongly interacting sector: ${\cal L}_{\rm int}\supset \tilde\lambda \bar f{\cal O}_f$. In the anarchic scenario there are no flavor symmetries in the new sector, such that all flavor transitions involving ${\cal O}_f$ are allowed and of the same order. We consider a modification of the anarchic paradigm introducing a horizontal global symmetry U(1)$_F$ in the strong sector, under which the operators can be charged. We assume that U(1)$_F$ is respected by the linear interactions, and assign to the elementary fermion $f$ the same charge as ${\cal O}_f$. In this case the linear coupling $\tilde\lambda$ are block diagonal, with different blocks associated to sectors with different charges.

At a low energy scale $\Lambda_{\rm IR}$ of order few TeV, the dynamics of the strong sector generates a mass gap and massive resonances, with the masses of the lowest level of composite states being $m_*\sim\Lambda_{\rm IR}$. Assuming that the strong sector has an approximate scale invariance, the running of the couplings is driven by the dimension of the corresponding operator, $\Delta_{{\cal O}_f}$, leading to: $\tilde\lambda_f(\Lambda_{\rm IR})\sim \tilde\lambda_f(\Lambda_{\rm UV})(\Lambda_{\rm IR}/\Lambda_{\rm UV})^{\Delta_{{\cal O}_f}-5/2}$, and generating a hierarchy of mixing for operators with different dimensions when $\Lambda_{\rm IR}\ll\Lambda_{\rm UV}$.

The model also features a spontaneous breaking of U(1)$_F$ in the composite sector, via a charged complex scalar operator ${\cal O}_\phi$ which is a singlet under the SM gauge symmetry. We normalise the charge of ${\cal O}_\phi$ to 1, $P_F{\cal O}_\phi={\cal O}_\phi$, and we use $p_f$ for the charges of the fermions, that are assumed to be integer numbers.

Since the Higgs is a composite state of the strong sector, two effects enter in the Yukawa interactions. First, if the fermionic operators are charged under U(1)$_F$, the Yukawa interactions require insertions of ${\cal O}_\phi$, leading to operators with higher dimensions than in partial compositeness. Second, the interactions with the elementary fermions are mediated by the linear mixing, such that the quark masses depend on the mixing of each chirality and on the number of insertions of ${\cal O}_\phi$.

We find it useful to describe the low energy limit of the above dynamics in terms of a two-site theory, with the elementary fermions and gauge fields of the SM associated to one site, and the first level of resonances of the strong sector, including the Higgs, associated to the other site~\cite{Contino:2006nn}. We will use small letters for the elementary fields and capital letters for the composite resonances. The resonances will have masses $m_*$, that we take of the same order, couplings $g_*$ that are taken as: $1< g_*\ll 4\pi$, and we define $f=m_*/g_*$. For simplicity we assume that there is only one fermionic resonance for each elementary fermion, and that they have the same quantum numbers under the SM gauge symmetry as the elementary ones. We also assume that there is only one complex scalar resonance excited by the operator ${\cal O}_\phi$. 

The interactions between elementary and composite fermions are given by:
\begin{equation}
  {\cal L}^{(\rm mix)} \supset -f(\lambda_{qj}\bar q_j Q_j + \lambda_{uj}\bar u_j U_j + \lambda_{dj}\bar d_j D_j) + {\rm h.c.} \ ,
  \label{eq-linearmix}
\end{equation}
where $\lambda_{fj}$ does not mix generations if $p_j\neq p_k$ for $j\neq k$. 
 
We find it useful to parametrise $\lambda_{fj}$ as:
\begin{equation}
\lambda_{fj}=g_*\epsilon_{fj}=g_*\lambda_C^{n_{fj}} \ , \qquad f=q,u,d \ ,
\end{equation}
with $\epsilon_{fj}$ being the degree of compositeness of the fermion $f_j$, and $\lambda_C\simeq 0.22$, such that we parametrise the mixing in powers of $\lambda_C$. Note that $n_{fj}$ do not need to be integer numbers.

We consider now the main interactions needed for our analysis. The strong sector has a global symmetry that contains the SM gauge symmetry group, as well as U(1)$_F$ factor. There are spin 1 resonances created by the conserved currents associated to these symmetries, that interact with the fermion resonances. Calling $F_\mu$ the lightest spin 1 resonance of U(1)$_F$ and $P_F$ its generator, the composite sector contains the interactions:
\begin{equation}
{\cal L}_{F}^{(\rm cp)} = g_{*F}\bar Q_j P_F \Fslash Q_j + g_{*F}\bar U_j P_F \Fslash U_j + g_{*F}\bar D_j P_F \Fslash D_j  \ ,
\end{equation}
that are not flavor universal if $P_F$ is not proportional to the identity. There are also interactions with the resonances associated to the SM gauge group that are flavor universal.

There are higher dimensional interactions with the Higgs, that require insertions of either $\Phi$ or $\Phi^\dagger$ to compensate the U(1)$_F$ charges of the fermions, with $\Phi$ being the effective complex scalar field describing the lowest lying excitation of ${\cal O}_\phi$: 
\begin{equation}
{\cal L}_{y}^{(\rm cp)} = - X^u_{jk}\bar Q_j \tilde H \left(\frac{\Phi^{(\dagger)}}{\Lambda}\right)^{\beta^u_{jk}}U_k - X^d_{jk}\bar Q_j H \left(\frac{\Phi^{(\dagger)}}{\Lambda}\right)^{\beta^d_{jk}}D_k  + {\rm h.c.}
\end{equation}
with $\Lambda$ the scale at which these operators are generated and:
\begin{equation}
\beta^u_{jk}=|p_{qj}-p_{uk}| \ , \qquad \beta^d_{jk}=|p_{qj}-p_{dk}| \ .
\end{equation}
The use of either $\Phi$ or $\Phi^{\dagger}$ depends on the sign of $(p_{qj}-p_{uk})$ and $(p_{qj}-p_{dk})$. We will consider an anarchic scenario, in which all the coefficients of the coupling $X_f$ are of the same order, $\mathcal{O}(g_*)$. The scale $\Lambda$ is expected to be of order $f$, although it depends on the ultra-violet dynamics that generates these interactions.

There is a spontaneous breaking of U(1)$_F$ as $\Phi$ acquires a vacuum expectation value: $\langle\Phi\rangle=\delta\Lambda$, with $\delta\lesssim 1$. When evaluating $\Phi$ in its vev, Higgs Yukawa couplings with the composite fermions are then generated:
\begin{equation}\label{eq-Yf}
Y^u_{jk} = X^u_{jk}\delta^{\beta^u_{jk}} \ , \qquad Y^d_{jk} = X^d_{jk}\delta^{\beta^d_{jk}} \ ,
\end{equation}
with no sum over repeated indices. We take the order of magnitude of the vev to be of the size of the Cabibbo angle in units of $\Lambda$: $\delta\sim\lambda_C$.

There can also be interactions with spin 0 composite states, similar to the Higgs interactions if the states are neutral under U(1)$_F$, and with a global shift of $\beta$ for charged states. The Yukawa couplings of these spin 0 states, called $\widetilde{X}^f$, are of $\mathcal{O}(g_*)$, and they are in general not aligned with $X^f$. We will consider the effect of these interactions in the next section.

\subsection{Fermion masses and mixing}
The mixing generates interactions between the elementary fermions and the Higgs. To leading order in $(v/f)$ and mixing the mass matrices of the light fermions of the SM are:
\begin{equation}
M^f_{j_k} \simeq v (\epsilon_q Y^f\epsilon_f)_{jk} = v\lambda_C^{n_{qj}}X^f_{jk}\delta^{\beta^f_{jk}}\lambda_C^{n_{fj}} = vX^f_{jk}\lambda_C^{n_{qj}+\beta^f_{jk}+n_{fk}}\ , \qquad f=u,d \ , \label{eq-massmatrix}
\end{equation}
where the last equality is obtained by taking $\delta\sim\lambda_C$. A diagonalization of this matrix will result in the spectrum of the physical states, along with the rotations between interaction eigenstates and mass eigenstates. For a given pattern of charges $\beta^{f}$ and mixings $\epsilon_f$, this system can be solved perturbatively in powers of $\lambda_C$. Moreover, the mass spectrum and rotation angles can be approximately found under suitable conditions. By considering a system with two generations, for $M_{11}\ll M_{12,21}$, the eigenvalues and eigenvectors can be estimated as:
\begin{align}
& m_{fj} \sim g_*v\lambda_C^{n_{qj}+\beta^f_{jj}+n_{fj}}\ ,  \label{eq-spectrum}\\
& \theta^f_{L,jk} \sim \lambda_C^{n_{qj}-n_{qk}+\beta^f_{jk}-\beta^f_{kk}} \ , \quad j<k \ , \label{eq-thetaLjk}\\
& \theta^f_{R,jk} \sim \lambda_C^{n_{fj}-n_{fk}+\beta^f_{kj}-\beta^f_{kk}} \ , \quad j<k \ . \label{eq-thetaRjk}
\end{align}
Although a proof of a similar formula for three generations is more involved, we have checked by performing a perturbative diagonalization in powers of $\lambda_C$, that for the cases considered in this paper these estimates work well when considering three generations. Thus we will make extensive use of these estimates.
In some cases we find corrections to these estimates, that end up being of the same order in powers of $\lambda_C$, we find this to be the case for quarks of the second generation.
This occurs when the Yukawa matrices have off-diagonal entries that contribute at the same order as the diagonal ones. 

\subsubsection{Solutions leading to the SM quark masses and mixing angles}
We consider the case in which the rotations into physical states $U^u_{L}$ and $U^d_{L}$ are of order $V_{\rm CKM}$. Since the most stringent constraints from flavor transitions arise from the $K$-system, we have explored the possibility to obtain $\theta^d_{L,jk}$ smaller than the CKM angles, but we have not found viable solutions of that kind. To minimise the flavor transitions we look for charge textures able to generate suppressed $\theta^d_{R,jk}$ and $\theta^u_{R,jk}$.

Avoiding suppression of the top mass requires that we take $n_{q3}=n_{u3}=\beta^u_{33}=0$, this in turn means equality of the third generation charges: $p_{q3}=p_{u3}$. For simplicity we pick: $p_{q3}=p_{u3}=0$; a nonzero value results in a shift of all charges by that value. By making use of Eq.~(\ref{eq-thetaLjk}), to reproduce the CKM angles in the up sector we demand:
\begin{align}
& |p_{q1}|\simeq 3-n_{q1} \ , \qquad |p_{q2}|\simeq 2-n_{q2}  \ , \label{eq-pq1pq2}\\
& n_{q1}-n_{q2} + |p_{q1}-p_{u2}| - |p_{q2}-p_{u2}|\simeq 1 \ ,
\end{align}
where we have assumed $n_{q1}\leq 3$ and $n_{q2}\leq 2$.
Since the same $n_{qj}$ and $p_{qj}$ enter in the down-sector, Eq.~(\ref{eq-pq1pq2}) leads to $\theta^d_{L,13}\sim\lambda_C^3$ and $\theta^d_{L,23}\sim\lambda_C^2$, {\it i.e.}: $(U^d_{L})_{j3}\sim (V_{\rm CKM})_{j3}$.
Notice that, although the charges of the fermions are integer numbers, the previous Eqs. do not require the degree of compositeness parametrised by $n_{fj}$ to be associated with an integer number, thus in the rest of the paper we will take $n_{fj}$ to be continuous variables.

For the bottom we take $p_{d3}=0$, therefore the ratio $m_b/m_t$ is controlled by $\epsilon_{d3}$, with $n_{d3}\sim 3$, see Eq.~(\ref{eq-mbovermt}). We will write the hierarchy of Right-handed down couplings in terms of $\epsilon_{d3}$, as $\epsilon_{d1,2}= \lambda_C^{n_{d1,2}} \epsilon_{d3}$, while remembering that this hierarchy is further suppressed with respect to the hierarchy of the up sector by a factor of $\lambda_C^3$, in the rest of the work we will make extensive use of this relation.
We can then write a solution, consisting of the following charges, and parametrised by the Right-handed hierarchies $n_{ui},\ n_{di}$:
\begin{align}
& p_{q1}=-1 \ , \quad  p_{q2}=0 \ , \quad n_{q1}=n_{q2}=2 \ , \nonumber \\
& p_{u1}\simeq -7+n_{u1},\quad p_{u2}\simeq  2-n_{u2},\quad n_{u1}\in[0,6],\quad n_{u2}\in[0,2]\ , \nonumber \\
& p_{d1}\simeq  -4+n_{d1},\quad p_{d2}\simeq 1-n_{d2},\quad n_{d1}\in[0,3],\quad n_{d2}\in[0,1] \ . \label{eq-charges}
\end{align}
The left charges are fixed once we pick $n_{q1}=n_{q2}=2$. There is freedom in the overall signs of the charges, but not in the relative sign between them. That is, in order to reproduce $U_{L}^{u \dagger} \simeq V_{\rm CKM}$, we make the choice $p_{q1}\leq 0$, $p_{q2}\geq 0$. The choice of $n_{q1}=n_{q2}=2$ can be explained by looking at the interaction of the Left-handed quarks with spin 1 resonances, as we describe in sec.~\ref{sec-intwres}. 
Once the Left-charges have been fixed, using Eq.~(\ref{eq-spectrum}) allows to fix the Right-charges, as a function of $n_{fi}$. The allowed ranges for the parameters $n_{fi}$ is thus limited by the same equation, such that larger $n_{fi}$ with fixed $n_{qi}$ and $\beta_{ii}$ implies a higher suppression of the fermion masses. 

For the set of solutions the $\beta$ matrices as a function of the parameters $n_{fi}$ are:
\begin{equation}
  \beta^u(n_{ui}) = \begin{pmatrix} 6-n_{u1} & 3-n_{u2} & 1 \\ 7-n_{u1} & 2- n_{u2} & 0 \\ 7-n_{u1} & 2-n_{u2} &  0 \end{pmatrix}, \quad\quad \beta^d(n_{di}) =   \begin{pmatrix} 3-n_{d1} & 2-n_{d2} & 1 \\ 4-n_{d1} & 1- n_{d2} & 0 \\ 4-n_{d1} & 1-n_{d2} &  0 \end{pmatrix} \ .
\end{equation}
Notice that when $n_{fi}=n_{fi}^{\rm max}$ (i.e.: their upper limits) the diagonal elements become zero and all the mass suppression comes from the coefficients $\epsilon_{fi}$, both Left and Right.

For the given solutions, the mass matrices of Eq.~(\ref{eq-massmatrix}) are independent of the parameters $n_{fi}$, thus the spectrum and the rotations matrices are independent of the values of these parameters. As already mentioned, a particularity of this solution with $n_{q1}=n_{q2}=2$ is that the rotation of the Left-handed down sector is also of order $U^d_{L} \simeq V_{\rm CKM}$. The same is not true for solutions that have $n_{q2}=1$, where we do find a suppression in the angle $\theta^d_{L,12}$, but those solutions have an enhancement of the non-diagonal Left-handed coupling of the up-sector with respect to APC, while not having a suppression in the non-diagonal Left-handed coupling of the down sector.

To lowest order in $\lambda_C$ the rotation matrices $U^f_R$ are given by:
\begin{equation}
  U^u_{R}\sim \begin{pmatrix}1&\lambda_C^5&\lambda_C^7\\ -\lambda_C^5 & 1 &\lambda_C^2 \\ -\lambda_C^7& -\lambda_C^2 & 1  \end{pmatrix}, \quad\quad\quad U^d_{R}\sim  \begin{pmatrix}1&\lambda_C^3&\lambda_C^4\\ -\lambda_C^3 & 1 &\lambda_C^1 \\ -\lambda_C^4& -\lambda_C^1 & 1  \end{pmatrix} 
\end{equation}
where we are omitting coefficients of $\mathcal{O}(1)$ that can be calculated in terms of the coefficients $X^f_{ij}$ of the Yukawa mass matrices. These coefficients, as well as higher order corrections that make these matrices orthogonal to a given non-trivial order in $\lambda_C$, are taken into account in our calculations.

Finally we need to address the charge degeneracy in these solutions. As one can see in Eq.~(\ref{eq-charges}), the charges of second and third generations of the Left-handed quarks are equal, as well as those of Right-handed down or up quarks when either $n_{d2}=n_{d2}^{\rm max}$ or $n_{u2}=n_{u2}^{\rm max}$. When charges are degenerated among different generations the couplings $\lambda_{fj}$ do not need to be diagonal in that subspace. One can choose a basis for the elementary fermions such that the coupling becomes triangular in the subspace~\cite{Panico-Pomarol,Frigerio-etal}, that is, there is a basis in which the mixing can be written as: 
\begin{equation}
  {\cal L}^{(\rm mix)} \supset - (\lambda_{q})_{jk} f\bar q_j Q_k - (\lambda_{u})_{jk} f\bar u_j U_k - (\lambda_{d})_{jk} f\bar d_j D_k + {\rm h.c.} \ ,
  \label{eq-linearmixT}
\end{equation}
with the coupling matrix
\begin{equation}
\lambda_f = \begin{pmatrix} \lambda_{f1} & 0 & 0 \\ 0 & \lambda_{f2} & 0 \\ 0 & t^f_{32} \lambda_{f2} & \lambda_{f3} \end{pmatrix}
\end{equation}
with $t^f_{32}=\mathcal{O}(1)$, in the case of generations 2 and 3 having degenerate charges. If another pair were degenerated, then $t^f_{ij}\neq 0,\ i > j$. This can further be parametrised as $\lambda_f = t^f g_* \epsilon_f$, with $t^f$ a lower-triangular matrix with its diagonal elements equal to 1. Thus the effect of the degeneracy is the insertion of the matrices $t^f$, with $f= q,\ u,\ d$, depending on which fermions are degenerate. As we will see below, this degeneracy will not play a significant role in {most of our analysis, as the leading order in $\lambda_C$ is dominated by the diagonal contributions or, in some particular cases, for this set of solutions the corrections are of the same order as the diagonal ones. $t^q$ does however play a role in the modified Left-handed coupling between $Ztc$, where the absence of this correction results in a higher suppression of the coupling for certain values of the parameters.

\subsection{Interactions with resonances}\label{sec-intwres}
In order to study the flavor constraints, we must first obtain the couplings between the elementary fermions and the resonances of the theory, that are responsible of mediating flavor transitions. We find it useful to distinguish between interactions with resonances of either spin 1 or 0. These interactions can be written in terms of the elementary-composite mixings, and the couplings between resonances. One must also rotate the elementary fermions into the mass eigenstates. We have, for spin 1 interactions: 
\begin{align}
& g^f_{L,ij}\simeq g_* (U^{f\dagger}_{L} \epsilon_q c_q \epsilon_q U^f_{L})_{ij} \ , \nonumber\\
& g^f_{R,ij}\simeq g_* (U^{f\dagger}_{R} \epsilon_f c_f \epsilon_f U^f_{R})_{ij} \ , \qquad f=u,d \ ,
\end{align}
where $c_q$ and $c_f$ are diagonal matrices with coefficients of ${\cal O}(1)$, that are not degenerate for the case of U(1)$_F$.

In the case of small mixing angles, we can approximate the rotations, and get an estimate for the different flavor transitions: 
\begin{align}
& g^f_{L,12}\sim g_*  [\theta^f_{L,12}(\epsilon_{q1}^2-\epsilon_{q2}^2)-\theta^f_{L,13}\theta^f_{L,23}\epsilon_{q3}^2] \ , \nonumber \\
& g^f_{L,23}\sim g_*  [\theta^f_{L,23}(\epsilon_{q2}^2-\epsilon_{q3}^2)-\theta^f_{L,13}\theta^f_{L,12}\epsilon_{q1}^2] \ , \nonumber \\
& g^f_{L,13}\sim g_*  [\theta^f_{L,13}(\epsilon_{q1}^2-\epsilon_{q3}^2)-\theta^f_{L,12}\theta^f_{L,23}\epsilon_{q2}^2] \ , \label{eq-gLij}
\end{align}
Equivalent expressions hold for the Right-handed couplings, by replacing $\epsilon_q \to \epsilon_f$, and $\theta^f_{L}\to \theta^f_{R}$. 

For the interaction with a spin 0 resonance, we have the following structure: 
\begin{align}
y^f_{ij}= (U^{f\dagger}_{L} \epsilon_q \widetilde{Y}^f \epsilon_f U^f_{R})_{ij}  \ ,
\end{align}
which involves the matrix $\widetilde{Y}^f$, which has the same structure as the Yukawa couplings with the Higgs, but is not necessarily aligned to it, and hence will not be diagonalized by rotations $U^f_{L,R}$. That is, $  \widetilde{Y}^f_{jk} = \widetilde{X}^f_{jk} \, \delta^{\beta^f_{jk}}$, with $\tilde{X}$ being $\mathcal{O}(g_*)$ coefficients.

Let us briefly discuss the structure of the flavor violating couplings. As can be seen from the interactions with spin-1 resonances in the small-angle approximation, Eq.~(\ref{eq-gLij}), flavor transitions depend on an interplay between the degree of compositeness and the mixing angles. On one hand in our model the degree of compositeness of some chiral fermions can be larger than in APC, in some cases increasing flavor violation, on the other hand the Right-handed mixing angles can be much smaller than in APC, their size being determined by the Froggatt-Nielsen charge of the fermions. Below we show that, given our choice of charges, the Right-handed mixing angles are very suppressed and the product is smaller than in APC, whereas the Left-handed couplings are of the same size as in APC. A similar situation holds for the couplings with spin 0 states, although in this case analytical expressions are larger and more complicated.

Let us evaluate these interactions in the case of the solution described above. We have seen how the parameters $n_{d1,d2}$ and $n_{u1,u2}$ are not fixed, but $n_{qi}$ are, as are the rotations into physical states. This means that out of the couplings above, we expect $g^f_{L}$ and $y^f$ to be independent of $n_{fi}$. For the spin 0 interactions this is a consequence of the shape of matrices $\beta^{f}$, as the combination $\epsilon_q  \cdot \lambda_C^{\beta^f}\cdot \epsilon_f$ produces a matrix that is independent of the $n_{fi}$. When evaluating these interactions, we must take into account that there may be phases in the different terms, and as such we must avoid artificial cancellations.

For interactions with a spin 1 resonance, we look at Eq.~(\ref{eq-gLij}) for the Left-handed coupling. Using that $n_{q1}=n_{q2}=2$, and $U^d_{L}\sim U^u_{L} \sim V_{\rm CKM}$, we have as the leading order:
\begin{equation}\label{eq-gLA}
g^d_L\sim g^u_L\sim g_* \epsilon_{q3}^2\begin{pmatrix} \lambda_C^4 & \lambda_C^5 & \lambda_C^3 \\ \dots & \lambda_C^4 & \lambda_C^2 \\ \dots & \dots & \lambda_C^0\end{pmatrix} \ ,
\end{equation}
where the lower triangular block is not shown since the matrix is symmetric. Notice that these flavor violating couplings are of the same size as in APC.

The Right-handed couplings of the spin 1 resonances depend on the values of $n_{fi}$, as $U^f_{R}$ are fixed all the dependence will be through the values of $\epsilon_{f}$. We get:
\begin{equation}\label{eq-gLB}
g^d_R\sim g_* \epsilon_{u3}^2\begin{pmatrix} \lambda_C^{6-12} & \lambda_C^{9-11} & \lambda_C^{10} \\ \dots & \lambda_C^{6-8} & \lambda_C^7 \\ \dots & \dots & \lambda_C^6\end{pmatrix} \ ,
\qquad
g^u_R\sim g_* \epsilon_{u3}^2\begin{pmatrix} \lambda_C^{0-12} & \lambda_C^{5-9} & \lambda_C^{7} \\ \dots & \lambda_C^{0-4} & \lambda_C^2 \\ \dots & \dots & \lambda_C^0\end{pmatrix} \ ,
\end{equation}
where again the lower triangular block is equal to the upper one. The coefficients that do not involve the third generation depend on $n_{fi}$ through some non-trivial functions, the range shown in the exponent of these coefficients correspond to the interval of those functions for the values of $n_{fi}$ in Eq.~(\ref{eq-charges}). On the other hand the coefficients involving the third generation are dominated by the contributions depending on $n_{u3}$ or $n_{d3}$, which are fixed in our solution. 
Approximate expressions for these functions can be obtained straightforwardly by using the small angle approximation.
To visualise this dependence we take $n_{fi}$ real, we evaluate the $\mathcal{O}(1)$ numerical coefficients taking care to avoid spurious cancellations and we obtain the contour plots of $\log_{\lambda_C}(g^f_{R,12}/g^{f,{\rm (APC)}}_{R,12})$ shown in Fig.~\ref{fig-gRRud}, namely the power of $\lambda_C$ that suppresses the coupling compared with APC. 
For down and up sectors a larger suppression requires larger values of $n_{fi}$, as well as $n_{d1}>0$ and $n_{u1}>1$. \footnote{Taking $n_{fi}$ to be integer, to have a suppression with respect to APC one has to choose $n_{u2}=n_{u2}^{\rm max}$ and $n_{d2}=n_{d2}^{\rm max}$. This causes the charges of the Right-handed second generation quarks to be degenerate with those of the third generation.}

\begin{figure}[H]
  \centering
  \begin{subfigure}[b]{0.47\textwidth}
    \centering
    \includegraphics[width=\textwidth]{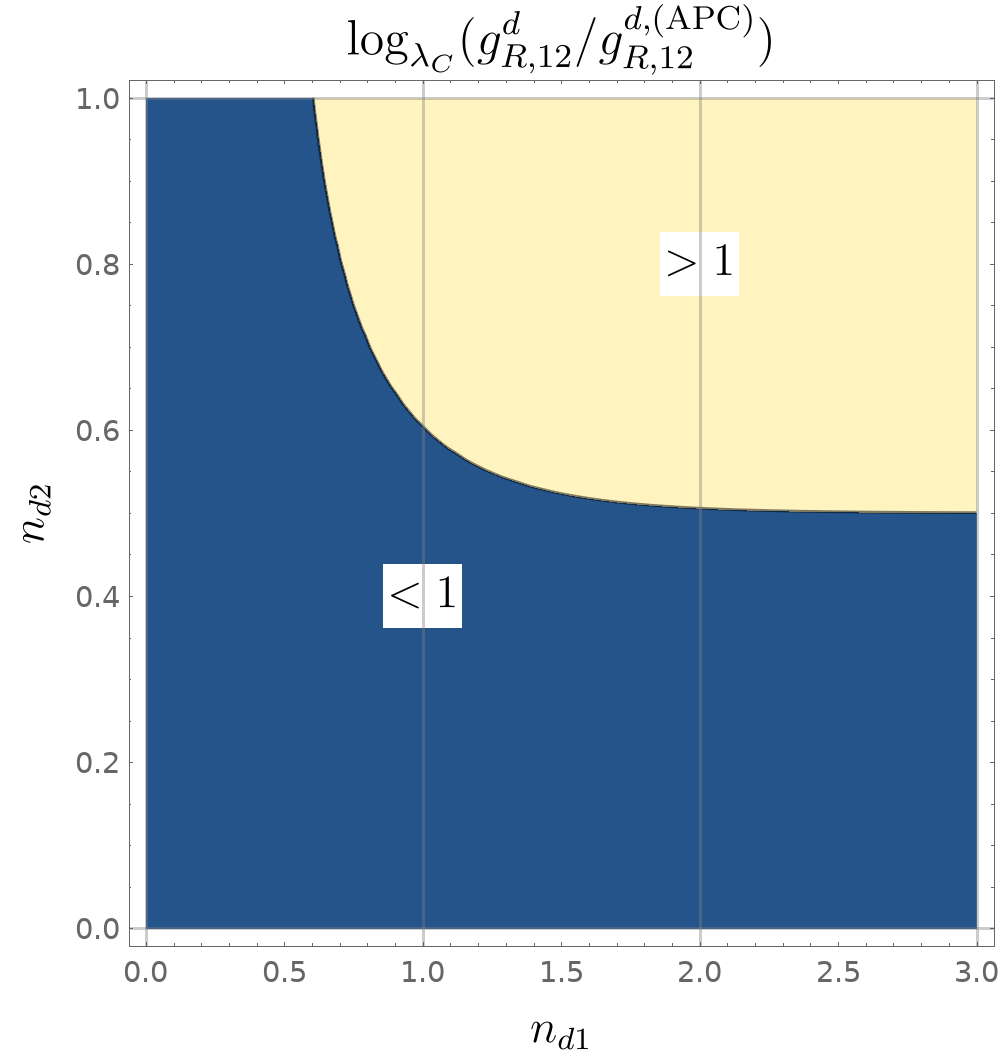}
  \end{subfigure}
  \hfill
  \begin{subfigure}[b]{0.47\textwidth}
         \centering
         \includegraphics[width=\textwidth]{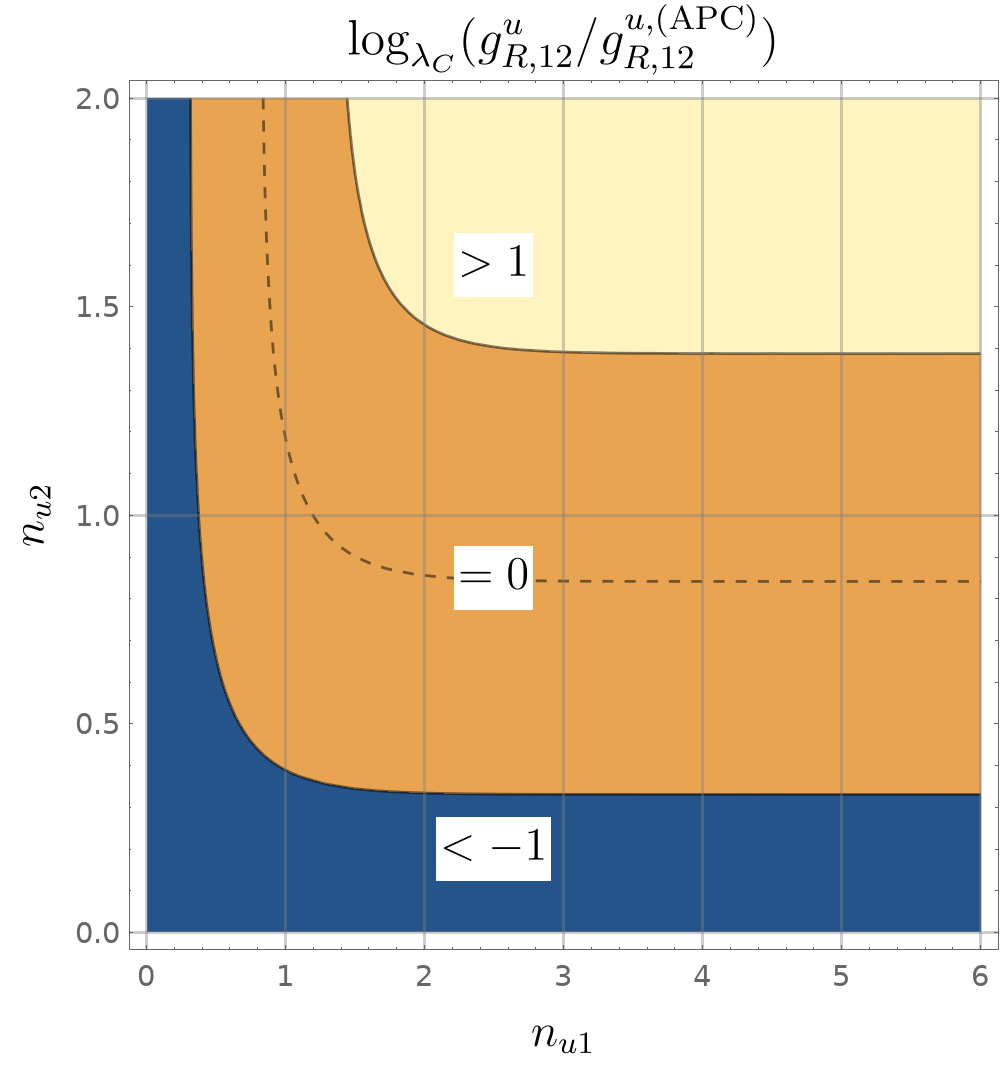}
  \end{subfigure}
  \caption{Flavor violating couplings of spin 1 resonances with light Right-handed quarks, compared with APC. On the left we show
$\log_{\lambda_C}(g^f_{R,12}/g^{f,{\rm (APC)}}_{R,12})$ for down-type quarks, and on the right for up-type quarks, as function of the degree of compositeness of the quarks parametrised by the exponents $n_{fi}$. Darker colour (left and down region) corresponds to smaller exponent, and lighter colour (up and right region) to larger exponent. For down (up) quarks, the exponent in the upper-right corner is 2 (2), and in the lower-left corner is 0 (-2).}
  \label{fig-gRRud}
\end{figure}

For the couplings with a spin 0 resonance we have to distinguish between up and down sectors. We get:
\begin{equation}\label{eq-gfLR}
g_{LR}^d \sim g_* \epsilon_{q3} \epsilon_{u3} \begin{pmatrix} \lambda_C^8 & \lambda_C^7 & \lambda_C^6 \\ \lambda_C^9 & \lambda_C^6 & \lambda_C^5 \\ \lambda_C^7 & \lambda_C^4 & \lambda_C^3\end{pmatrix} ,\quad\quad\quad g_{LR}^u \sim g_* \epsilon_{q3} \epsilon_{u3} \begin{pmatrix} \lambda_C^8 & \lambda_C^5 & \lambda_C^3 \\ \lambda_C^9 & \lambda_C^4 & \lambda_C^2 \\ \lambda_C^7 & \lambda_C^2 & \lambda_C^0\end{pmatrix} \ .
\end{equation}
Here we show only the dominant term in powers of the Cabibbo angle, and as stated above, these flavor transitions are all independent of the values of $n_{fi}$. 

As mentioned above, in the presence of degenerate charges one must introduce couplings between elementary and composite sectors which are non-diagonal, but triangular, with nonzero elements connecting these generations. We considered the effect of these non-diagonal couplings, present in the left sector as $n_{q2}=2$, and in the right sector only if $n_{d2}=n^{\rm max}_{d2}$ or $n_{u2}=n^{\rm max}_{u2}$. This is easily calculated by using the appropriate insertions of $t^f$ in our calculations. We found that the leading order in powers of $\lambda_C$ is unchanged in interactions with either scalar or vector resonances, this is a particular result of the solution that we are considering, in which the new contribution is of the same order as those arising from diagonal $\lambda_f$.

\subsection{Coefficient of dipole operators}
We find it useful to discuss in this section the general flavor structure of the Wilson coefficient of dipole operators, that are induced at one loop level by exchange of resonances. The dipole operators can be of different kinds, depending on which boson closes the loop. In Eq.~(\ref{eq-dipoles}) we show an estimate in terms of the relevant parameters, in which we add different contributions that include: U(1)$_F$ resonance in the first three terms, neutral Higgs in the fourth term, and charged Higgs in the last term,
\begin{align}\label{eq-dipoles}
d^f_{ij}=g_* [U^{f\dagger}_{L} \epsilon_q (P_q^2Y^f+P_qY^fP_f+Y^fP_f^2+Y^fY^{f\dagger} Y^f+Y^{f'}Y^{f'\dagger} Y^f) \epsilon_f U^{f}_{R}]_{ij}  \ . 
\end{align}  
where $f$ is either $u$ or $d$, and $f' \neq f$.
There are also contributions with exchange of spin 1 resonances with the quantum numbers of the gluons, $W$'s and $Z$ that in general are suppressed~\cite{Neubert}, eventually we do not expect them to be larger than that of $F$.

Dipole operators can involve either $\bar q_{L,i}$ and $q_{R,j}$, or $\bar q_{R,i}$ and $q_{L,j}$, as usual in the case $i\neq j$ we call their Wilson coefficients $C_{ij}$ and $C_{ij}'$

\section{Constraints}\label{sec-constraints}
In this section we study the main flavor constraints present in our model for the specific solution defined above. 
We also analyse constraints from dijets at LHC and from mixing of the real component of $\Phi$ with the Higgs field.

We make use of the couplings calculated in the previous section, at leading order in powers of $\lambda_C$. We compare the size of the Wilson coefficients with respect to those of APC. 
We consider several flavor and CP violating processes, that are conveniently organised as: $\Delta F=2$, $\Delta F=1$ and $\Delta F=0$.  

\subsection{Constraints from $\Delta F=2$ transitions}
These processes come from 4-fermion contact interactions that, by using Fierz identities, can be reduced to eight dimension six operators whose Wilson coefficients have stringent bounds~\cite{07070636}. We follow the notation of that reference:
\begin{equation}
{\cal H}_{\rm eff}^{\Delta F=2} =\sum_{i,j}\left(\sum_{k=1}^5C^{q_iq_j}_k Q^{q_iq_j}_k+\sum_{k=1}^3\tilde C^{q_iq_j}_k \tilde Q^{q_iq_j}_k\right)
\end{equation}
with 
\begin{align}
& Q_1^{ij}=(\bar q_{L,j}^\alpha\gamma_\mu q_{L,i}^\alpha)(\bar q_{L,j}^\beta\gamma^\mu q_{L,i}^\beta) \ , \nonumber\\
& Q_2^{ij}=(\bar q_{R,j}^\alpha q_{L,i}^\alpha)(\bar q_{R,j}^\beta q_{L,i}^\beta) \ , \nonumber\\
& Q_3^{ij}=(\bar q_{R,j}^\alpha q_{L,i}^\beta)(\bar q_{R,j}^\beta q_{L,i}^\alpha) \ , \nonumber\\
& Q_4^{ij}=(\bar q_{R,j}^\alpha q_{L,i}^\alpha)(\bar q_{L,j}^\beta q_{R,i}^\beta) \ , \nonumber\\
& Q_5^{ij}=(\bar q_{R,j}^\alpha q_{L,i}^\beta)(\bar q_{L,j}^\beta q_{R,i}^\alpha) \ , 
\end{align}
where $\alpha$ and $\beta$ are color indices.
$\widetilde Q^{q_iq_j}_k$ can be obtained from $Q^{q_iq_j}_k$ by flipping the quark chiralities.

In APC the most stringent bounds for $\Delta F=2$ processes arise from $C_2^{sd}$, $\tilde C_2^{sd}$ and $C_1^{d_id_j}$, that require $m_*\gtrsim 5-8$~TeV, and from $C_4^{sd}$ that requires $m_*\gtrsim 10-20$~TeV, with some dependence on the details of the model.
Less stringent but still sizeable are the bounds from $C_{2,4}^{bd}$, $C_{2,4}^{bs}$, $C_{1,2,4}^{cu}$, that require $m_*\gtrsim 0.5-3$~TeV.~\cite{Isidori:2010kg,Panico:2015jxa}

Exchange of spin 1 and spin 0 resonances give contributions to these Wilson coefficients that can be written in terms of the couplings of the previous section. The spin 1 contributions are: 
\begin{align}
C_1^{ij} = \frac{(g^f_{L,ij})^2}{m_*^2} \ , \qquad 
\widetilde{C}_1^{ij}  = \frac{(g^f_{R,ij})^2}{m_*^2} \ , \qquad 
C_4^{(1),ij}  = \frac{g^f_{L,ij} g^f_{R,ij}}{m_*^2} \ .
\end{align}
Spin 0 resonances contribute to the following coefficients:
\begin{align}
C_2^{ij} = \frac{(g^f_{LR,ij})^2}{m_*^2} \ , \qquad 
\widetilde{C}_2^{ij}  = \frac{(g^f_{LR,ji})^2}{m_*^2} \ , \qquad 
C_4^{(0),ij}  = \frac{g^f_{LR,ij} g^f_{LR,ji}}{m_*^2} \ .
\end{align}

To obtain a suppression of $C_1^{sd}$ and $C_1^{cu}$ in $K$ and $D$ systems as efficient as in APC requires $g_{f_L,12}\lesssim \lambda_C^5$. From the first term of Eq.~(\ref{eq-gLij}), assuming $n_{q1}\geq n_{q2}$ and $\theta^f_{L,12}\sim \lambda_C$ gives: $n_{q2}\geq 2$, whilst the second term is saturated for $\theta^f_{L,23}\theta^f_{L,13}\sim \lambda_C^5$, {\it i.e.}: when $\theta^f_{L,23}$ and $\theta^f_{L,13}$ are of CKM order. This is the reason why we have not considered solutions with $n_{q2}<2$. If $\theta^{\rm CKM}_{23}$ and $\theta^{\rm CKM}_{13}$ were generated one in the up- and the other in the down-sector, one could look for a solution with a suppressed second term in Eq.~(\ref{eq-gLij}), but we have not found such a solution.

We show the prediction for the Wilson coefficients normalised in terms of what is obtained for APC, we express this ratio in powers of $\lambda_C$ and focus on the exponent, such that larger exponent corresponds to larger suppression. This is shown in Table~\ref{table-df2}, where we write $\log_{\lambda_C} \left( C_X/C_X^{(\rm APC)}\right)$ for each meson. 

\begin{table}[H]
  \centering
  \def\arraystretch{1.3}%
\begin{tabular}{|c | c |  c | c|c|c|c|} 
  \hline 
 $\Delta F=2$ & $C_1$ &$\widetilde{C}_1$& $C_4^{(1)}$ & $C_2$ &$\widetilde{C}_2$ & $C_4^{(0)}$ \\
  \hline 
 $K$   & 0 & 0 to 4 & 0 to 2 & 0 & 4 &2  \\ [0.5ex] 
 \hline
 $B_d$ & 0 & 4 & 2 & 0  & 4 & 2  \\ 
 \hline
 $B_s$ & 0 & 0 & 0  & 0 & 0  & 0 \\
 \hline
 $D$   & 0 & -4 to 4 & -2 to 2 & 0 & 4 & 2 \\
 \hline
\end{tabular}
\caption{Summary of model results for the Wilson coefficients of $\Delta F = 2$ operators. Each cell contains the value of $\log_{\lambda_C} \left( C_X/C_X^{\rm (APC)}\right)$, that shows the amount of suppression (or enhancement) with respect to APC, for each meson. For an explanation of the meaning of cells with a range of values read the text.}
\label{table-df2}
\end{table}

Several patterns emerge in table~\ref{table-df2}. Columns corresponding to $C_1$ and $C_2$ are all zero, meaning that these coefficients are of the same size as in APC for the solution that we have chosen. The row corresponding to $B_s$ is also zero, thus having no suppression with respect to APC for this meson. As we have previously shown, for our model the left handed couplings with a spin 1 resonance are all of the same order as those in APC, thus $C_1$ has this same behaviour. $C_2$ is also of the same order as in APC, due to the fact that the couplings $g_{LR}^{f}$, Eq.~(\ref{eq-gfLR}), have the upper triangular block of the same order as APC. Regarding $B_s$, it also has couplings $g^d_{LR,32}$ and $g^d_{R,32}$ of the same size as APC, thus none of its Wilson Coefficients are suppressed. Regarding the dependence of these coefficients on $n_{u,d}$, as we discussed in the previous section, only those quantities that depend on $g^f_{R}$ can have a dependence on $n_{f}$, then only coefficients $\widetilde{C}_1$ and $C_4^{(1)}$ have a dependence with these parameters. This dependence can be understood from the plots in Fig.~\ref{fig-gRRud}, as those Wilson coefficients are proportional $(g^f_{R})^2$ and $g^f_{R}$, respectively. In the table we show the range of values that are obtained for $n_{f}$ varying in the corresponding intervals. For the $K$-meson a suppression is obtained as long as $n_{d1}$ and $n_{d2}$ are positive, with a suppression ${\Order}(\lambda_C^2)$ in $\widetilde C_1$ and ${\Order}(\lambda_C)$ in $C_4$ for $n_{d1}\gtrsim 1$ and $n_{d2}\gtrsim 0.5$, and a maximum suppression ${\Order}(\lambda_C^4)$ in $\widetilde C_1$ and ${\Order}(\lambda_C^2)$ in $C_4$ for the maximum values of $n_{d1,2}$. For the $D$-meson suppression is obtained for $n_{u1},n_{u2}\gtrsim 1$, with a suppression ${\Order}(\lambda_C^2)$ in $\widetilde C_1$ and ${\Order}(\lambda_C)$ in $C_4$ for $n_{u1},n_{u2}\gtrsim 1.5$, and a maximum suppression ${\Order}(\lambda_C^4)$ in $\widetilde C_1$ and ${\Order}(\lambda_C^2)$  in $C_4$ for the maximum values of $n_{u1,2}$.~\footnote{If only integers $n_f$ are considered and the ${\cal O}(1)$ parameters of the theory are strictly fixed to unity, one gets only the extremes of the interval shown in the cells in the case of the $K$-meson, with the suppression corresponding to $n_{d2}$ maximum and $n_{d1}>0$. In the case of the $D$-meson, for integer $n_f$, there are three different values for these Wilson coefficients depending on the value of $n_{ui}$, for $\widetilde C_1$: $-4,0,4$ and for $\widetilde C_4^{(1)}$: $-2,0,2$.}

Overall, we obtain that the most stringent constraints, that involve Right-handed couplings, can be suppressed, whereas the Left-handed ones are as in APC. This implies that a smaller Left-compositeness of the top quark $\epsilon_{q3}$ is required, trading it for larger Right-handed compositeness in order to accurately reproduce the top mass, as $\epsilon_{q3}= 1/g_*$ and $\epsilon_{u3}=1$.

\subsection{Constraints from $\Delta F=1$ transitions}
The main effects arise from the following classes of operators: dipole operators, that give the most stringent bounds of this subsection; penguin operators, that modify the $Z$ couplings and operators that modify the $W$ couplings. The last two produce smaller effects, and are rather model dependent, since by the introduction of symmetries some of them can be protected~\cite{Agashe:2006at}. Below we will focus on dipole operators as well as on flavor violating $Z$ couplings with the top.

\subsubsection{Dipole operators}
For $\Delta F=1$ processes the effective Hamiltonian with dipole operators can be written as~\cite{Neubert}:
\begin{equation}
{\cal H}_{\rm eff}^{\Delta F=1}=\sum_{i,j}(C^{q_iq_j}_7Q^{q_iq_j}_7+C'^{q_iq_j}_7Q'^{q_iq_j}_7) \ ,
\end{equation}
with
\begin{align}
Q^{q_iq_j}_{7,\gamma}=\frac{em_{q_i}}{16\pi^2}(\bar q_{Lj}\sigma^{\mu\nu}q_{Ri})F_{\mu\nu} \ ,
\qquad
Q^{q_iq_j}_{7,g}=\frac{g_sm_{q_i}}{16\pi^2}(\bar q_{Lj}\sigma^{\mu\nu}T^aq_{Ri})G^a_{\mu\nu} \ , \nonumber \\
Q'^{q_iq_j}_{7,\gamma}=\frac{em_{q_i}}{16\pi^2}(\bar q_{Rj}\sigma^{\mu\nu}q_{Li})F_{\mu\nu} \ ,
\qquad
Q^{q_iq_j}_{7,g}=\frac{g_sm_{q_i}}{16\pi^2}(\bar q_{Rj}\sigma^{\mu\nu}T^aq_{Li})G^a_{\mu\nu} \ .
\end{align}
with $T^a$ the color generators, $F_{\mu\nu}$ and $G_{\mu\nu}$ the field strength of photons and gluons.
The dominant bounds in APC arise from the Wilson coefficients of: $C_{7,g}^{(')sd}$, $C_{7,V}^{(')bs}$ and $C_{7,g}^{(')cu}$, that require $f\gtrsim 1-2$~TeV~\cite{Agashe:2008uz,Neubert,Panico:2015jxa}.

The dipole operators can be of different kinds, depending on which boson closes the loop. In Eq.~(\ref{eq-dipoles}) we added the different contributions, which include U(1)$_F$ resonances, neutral Higgs and charged Higgs in the loop. We summarise our results in table~\ref{table-df1}, where we show $\log_{\lambda_C}\left(d_{ij}^f/d_{ij}^{f,(\rm{APC})}\right)$ for the different flavor combinations involved.

\begin{table}[H]
  \centering
\def\arraystretch{1.3}%
\begin{tabular}{| c | c |  c | c|} 
 \hline 
 $\Delta F= 1$ & U(1)$_F$ &$(Y^f)^3$& $(Y^{f'})^2 Y^f$ \\
 \hline 
   $C_7(b s \gamma)$  & 2 & 0  & 0  \\
 \hline
   $C_7^{'}(b s \gamma)$  & $\begin{cases} 0&\ (n_{d2}=0)\\  6&\ (n_{d2}=1)  \end{cases}$& 0  & 0 \\
 \hline
 $C(s d g)$  &  0  & 0  & 0 \\
 \hline
 $C^{'}(s d g)$  &   2  & 2   &  2  \\
 \hline
 $C(c u g)$  &  0 &  0   &  0   \\
 \hline
 $C^{'}(c u g)$  &   2   & 2    &  2  \\
 
 \hline
 \end{tabular}
\caption{Summary of $\Delta F=1$ coefficients of dipole operators. We show the value of $\log_{\lambda_C}(C/C^{(\rm APC)})$, for the ratios of the Wilson coefficients normalised with respect to APC.}
\label{table-df1}
\end{table}

We obtain that in the case of $\Delta F = 1$, for most flavor transitions there is at least one contribution that is of the same size as in APC. This is not the case for one of the chiral structures of $sdg$ and $cug$, where we obtain a suppression $\lambda_C^2$. For $C_7^{'}(b s \gamma)$ we obtain a range of values that depend on $n_{d2}$.

As a summary, adding all the contributions and chiral structures we do not obtain suppressions compared with APC.

\subsubsection{$Ztc$ and $Zbs$ interactions}
Experimental bounds on ${\rm BR}(t\to Zc)$ lead to $g^Z_{tc}\lesssim 0.01$ for Left- and Right-handed chiralities~\cite{1702.01404,1803.09923}. Although these interactions are model dependent, since it is possible to protect one of the couplings with discrete symmetries~\cite{Agashe:2006at}, in general it is not possible to protect both chiral couplings at the same time. In partial compositeness one can estimate: $\delta g^Z_{tc}\sim (g/2) \epsilon_2\epsilon_3 c_{23}c_{33}(v/f)^2$, with $c_{ij}$ numbers that depend on the flavor structure of the model, $c_{23}$ connecting second and third generations in the composite sector, $\epsilon_j$ being the Left-handed (Right-) degree of compositeness for $\delta g^Z_{L,tc}$ ($\delta g^Z_{R,tc}$).

In the present model the coefficients $c_{ij}$ are given by $Y^u_{ij}/g_*\sim \lambda_C^{\beta^u_{ij}}$, Eq.~(\ref{eq-Yf}), and the degree of compositeness is given by $\epsilon_{fj}\sim\lambda_C^{n_{fj}}$. For the solutions of Eq.~(\ref{eq-charges}) we find that the Right-handed coupling has a size $\delta g^Z_{R,tc}\sim \lambda_C^2(v/f)^2$, that is of the same order as APC~\cite{1408.4525}. For $(v/f)^2\sim 0.1$, $\delta g^Z_{R,tc}\sim 5\times 10^{-3}$. The Left-handed coupling, however, has a dependence on parameters $n_{di}$. We find this coupling to be within the range $\lambda_C^2 (v/f)^2$ (for $n_{u2}=2$) to $\lambda_C^4 (v/f)^2$ ($n_{u2}<2$). We find that this coupling suffers from the effect of matrix $t^q$ present due to charge degeneracy between 2nd and 3rd generation Left-handed quarks. Without the effect of this matrix, however, coupling can be as small as $\lambda_C^6 (v/f)^2$ for $n_{u2}=0,\ n_{u1}<6$.

In the 14 TeV run of LHC, with 3000~fb$^{-1}$, it is expected to probe $\delta g^Z_{tc}\sim 3-6\times 10^{-3}$ at 95\%CL, testing the Right handed coupling of the present model, as well as APC, for $(v/f)^2\sim 0.1$.

For the $Zbs$ interaction, we calculate the Left-handed coupling $\delta g^Z_{L,bs}$ to be within the range $\lambda_C^2(v/f)^2$ (for $n_{d2}=1$) to $\lambda_C^4(v/f)^2$ (for $n_{d2}=0$). The Right-handed coupling has a size $\delta g^Z_{R,bs} \sim \lambda_C^7 (v/f)^2$, that is of the same order as APC.

$B_s\to \mu\mu$ strongly constraints $g^Z_{bs}$. By making use of the experimental measurements \cite{1307.5025,1307.5024,Buchalla}, for a generic spectrum of resonances, a crude estimate leads to $\epsilon_{q3}^2(v/f)^2\lambda_C^{2-4}\lesssim 10^{-5}$. This bound could be satisfied for $n_{d2}^{\rm min}$, $(v/f)^2=0.1$ and $\epsilon_{q3}\simeq 0.3$, but would require a tuning of order $\lambda_C^{-2}$ for $n_{d2}^{\rm max}$. The presence of symmetries protecting down-type $Z$ couplings can relax this tension. 

\subsection{Constraints from $\Delta F=0$ processes}
For $\Delta F =0$ we also look at the dipole operators, that in APC are dominated by the down-quark, that requires $f\gtrsim 4-5$~TeV~\cite{Neubert,Panico:2015jxa}.
We focus on $d$, $u$ and $c$ quarks. In order to get the leading contribution to these Wilson coefficients, we must analyse the misalignment taking place between the dipole operators and the mass matrix. This is of particular importance in the case of the U(1)$_F$ vector closing the loop, because although the matrix combinations in the first three terms of Eq.~(\ref{eq-dipoles}) are proportional to the same Yukawa matrix, insertions of the fermion charges $\alpha_q$ and $\alpha_f$ break the alignment at higher orders, thus contributing to a phase. 
In Table~\ref{table-df0} we show the power of the leading order contribution to these coefficients. 
We find a suppression of order $\lambda_C^2$ for the U(1)$_F$ in the loop, in all three quarks $d,\ u,$ and $c$.
When looking at contributions from the Higgs in the loop, with a cubic dependence on the Yukawa, no such alignment is present.
In those contributions we find, for both up and down sectors, a dependence on the parameters of our model, $n_{u1,u2}$ or $n_{d1,d2}$.
\begin{table}[H]
  \centering
\def\arraystretch{1.3}%
\begin{tabular}{| c | c |  c | c|} 
 \hline
 $\Delta F= 0$ & U(1)$_F$ & $(Y^f)^3$& $(Y^{f'})^2 Y^f$ \\
 \hline
 $ddV$& 2 &  $\begin{cases} & 2-6\ (n_{d1}<3) \\ & 0\ (n_{d1}=3) \end{cases}$  &  $\begin{cases} & 2\ (n_{u1}<6) \\ & 0\ (n_{u1}=6) \end{cases}$  \\
 \hline
$uuV$ & 2 & $\begin{cases} & 2-8\ (n_{u1}<6) \\ & 0\ (n_{u1}=6) \end{cases}$ &  $\begin{cases} & 2\ (n_{d1}<3) \\ & 0\ (n_{d1}=3) \end{cases}$  \\
 \hline
$ccV$ & 2 & $\begin{cases} & 2-4\ (n_{u2}<2) \\ & 0\ (n_{u2}=2) \end{cases}$ & 0  \\
 \hline
\end{tabular}
\caption{Summary of $\Delta F=0$ coefficients of dipole operators. We show the value of $\log_{\lambda_C}(C/C^{(\rm APC)})$ for the ratios of the Wilson coefficients normalised with respect to APC. For the U(1)$_F$ vector loop, as the coefficient can be aligned at zeroth order with the mass matrix, we show the first non-zero contribution.}
\label{table-df0}
\end{table}

To better understand the dependence of $\log_{\lambda_C}\left(d^f_{ii}/d^{f,\rm{(APC)}}_{ii}\right)$ on the parameters of the theory we show in Fig.~\ref{fig-Y3dip} the contribution from the neutral Higgs.
We see that for $ddV$ (left panel), no such suppression is found if $n_{d1}=n_{d1}^{\rm max}=3$, however, for smaller values of $n_{d1}$ the suppression can be of order $\lambda_C^2$ and larger. In the case of the up sector, we see a different dependence for the up quark (middle panel) or the charm quark (right panel). Whereas the contribution to the  up quark is of order APC for $n_{u1}=n_{u1}^{\rm max}=6$, and has larger suppressions for smaller values of $n_{u1}$, the contribution to the charm has a stronger dependence on $n_{u2}$, being of order APC when $n_{u2}=n_{u2}^{\rm max}=2$ and smaller for $n_{u2}< n_{u2}^{\rm max}$. Here we see a small tension with $\Delta F=2$ $D$-meson constraints that prefer larger $n_{u2}=n_{u2}^{\rm max}$ to suppress $\widetilde{C}_1$ and $C_4$, although there is a window where coefficients with Right-handed quarks are suppressed for both processes.

\begin{figure}[H]
  \centering
  \begin{subfigure}[b]{0.31\textwidth}
    \centering
    \includegraphics[width=\textwidth]{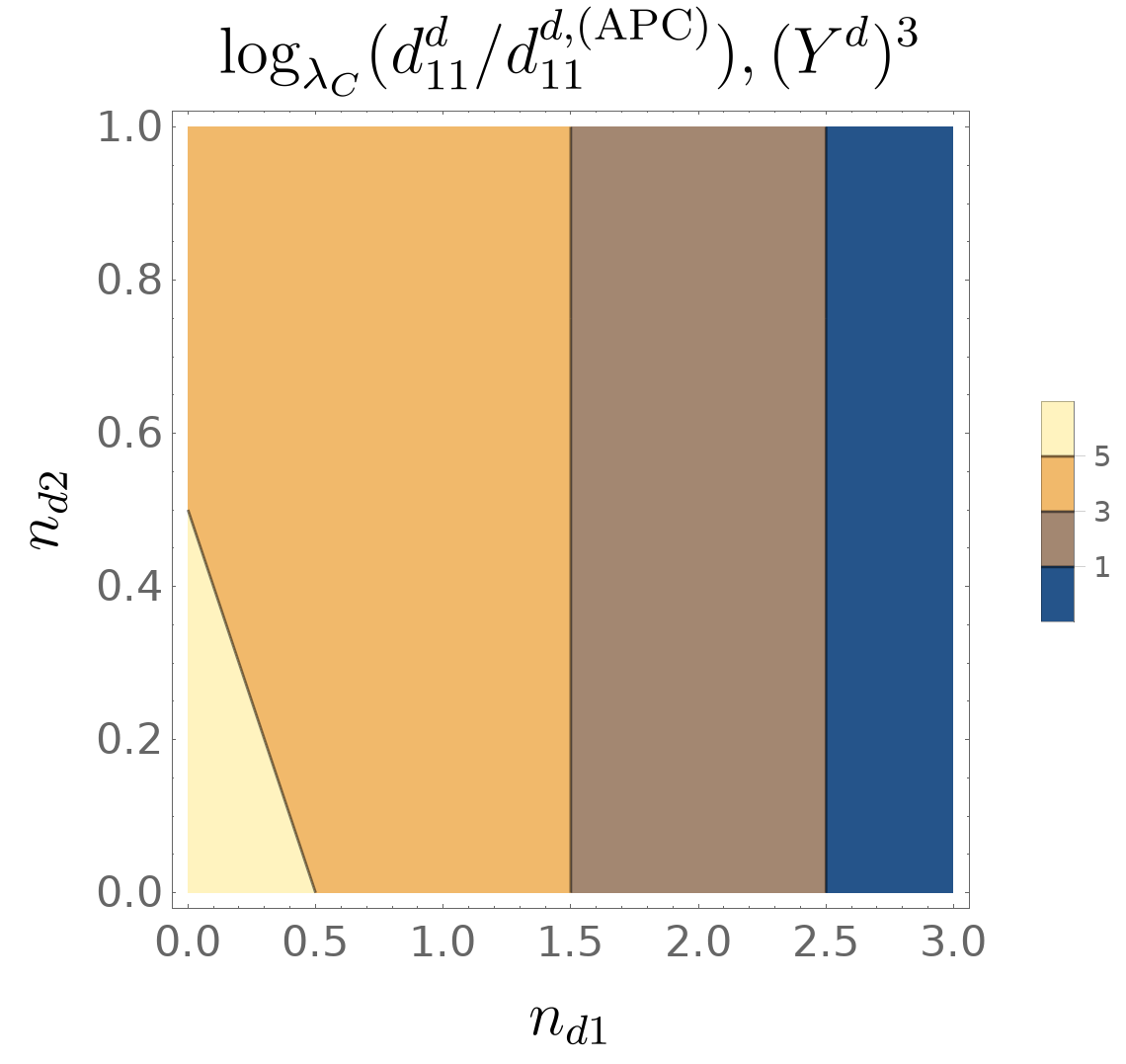}
    \caption{$ddV$}
  \end{subfigure}
  \hfill
  \begin{subfigure}[b]{0.31\textwidth}
         \centering
         \includegraphics[width=\textwidth]{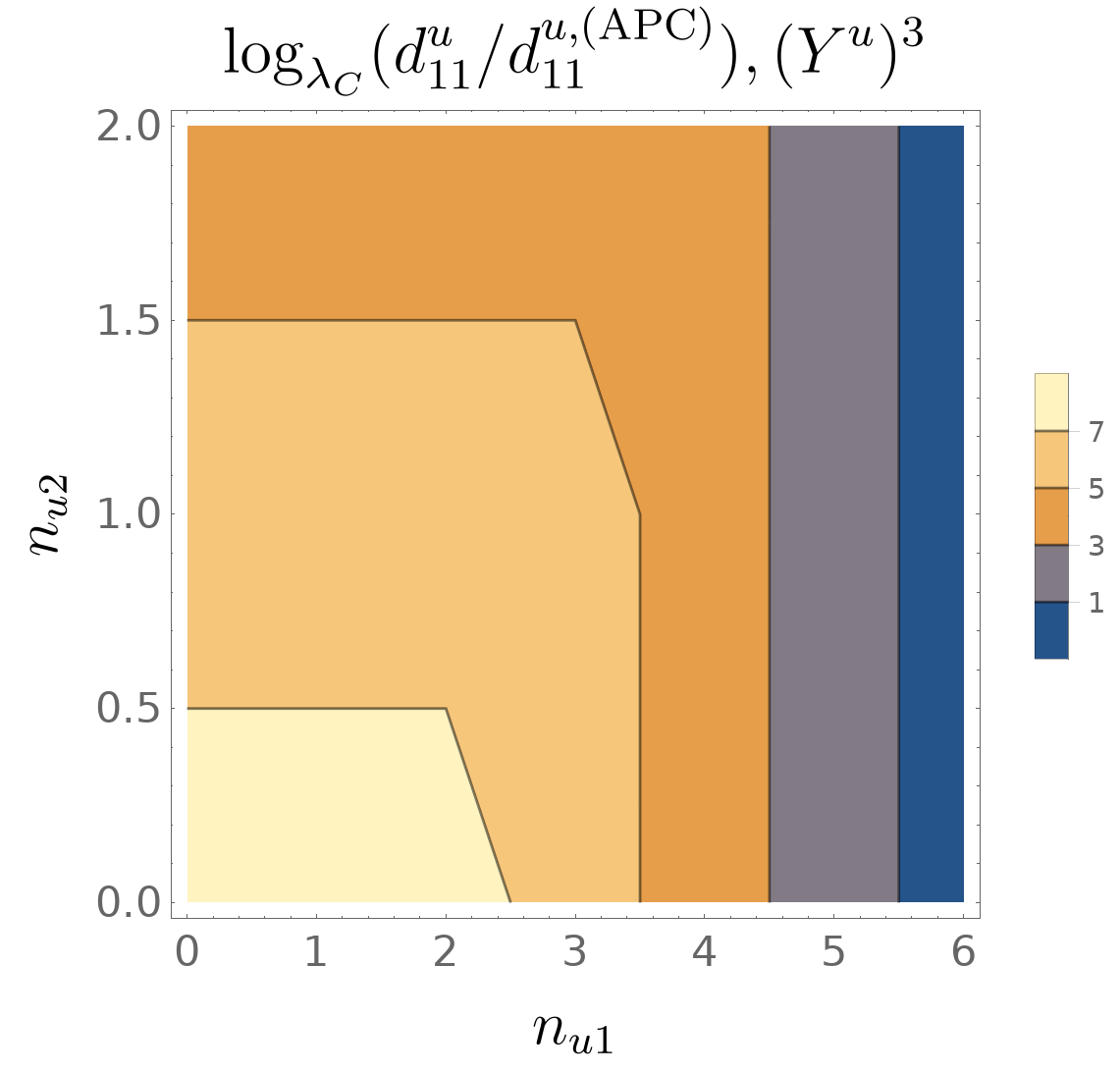}
         \caption{$uuV$}
  \end{subfigure}
  \hfill
  \begin{subfigure}[b]{0.31\textwidth}
         \centering
         \includegraphics[width=\textwidth]{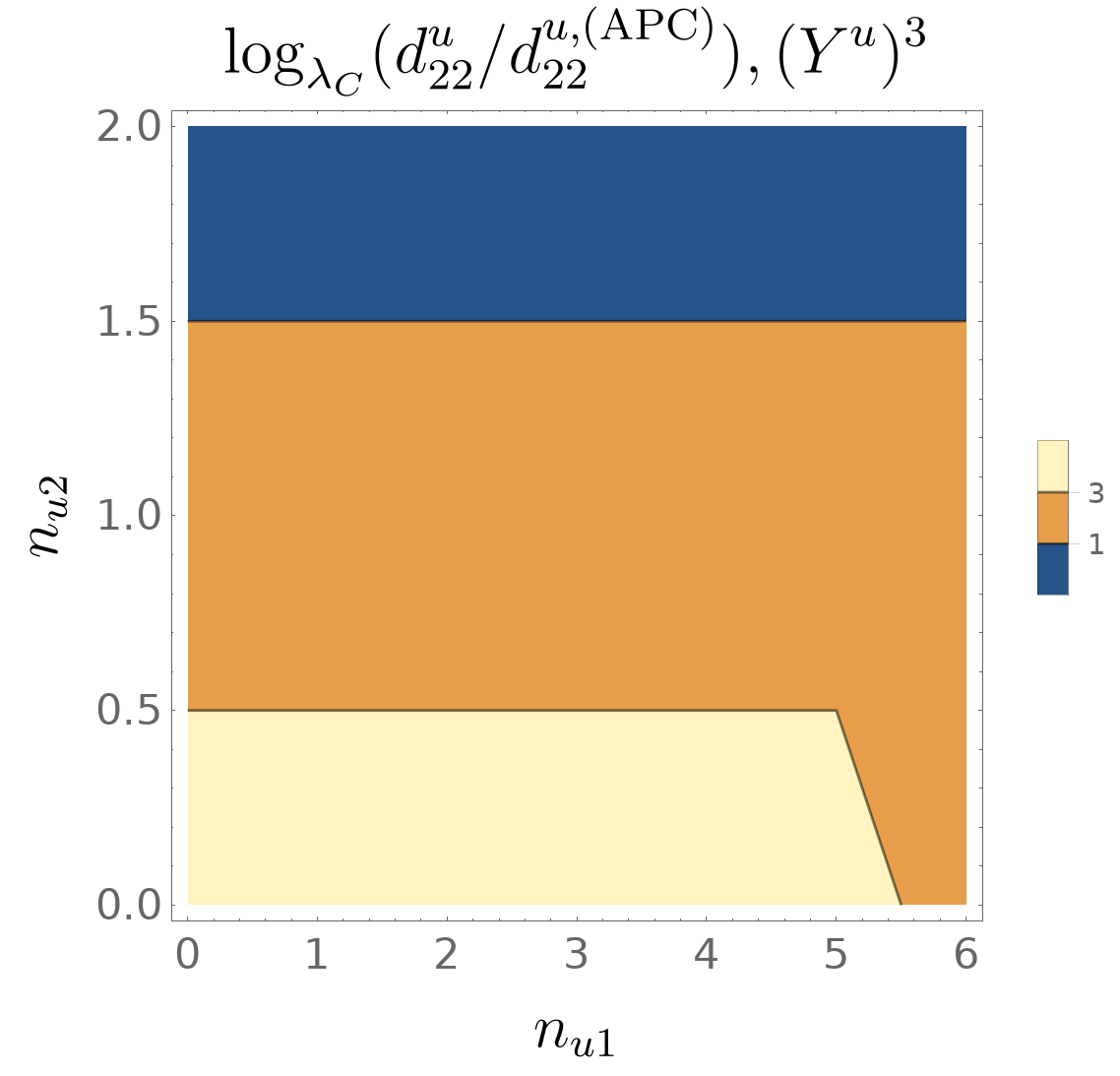}
         \caption{$ccV$}
  \end{subfigure}
  \hfill
  \caption{Wilson coefficient of flavor diagonal dipole operators compared with APC: $\log_{\lambda_C}\left(d^f_{ii}/d^{f,\rm{(APC)}}_{ii}\right)$ with neutral Higgs contributions, for down (left panel), up (middle panel) and charm (right panel) quarks, as function of $n_{fi}$ parameters of the theory. A darker colour corresponds to smaller exponent, and lighter colour to a larger exponent. In all three cases, the exponent in blue regions is $0$, thus leading to no suppression.}
    \label{fig-Y3dip}
\end{figure}

In the case of the charm dipoles, the contribution of a charged Higgs has the same size as in APC, irrespective of values of $n_{ui,di}$. The fact that no suppression is observed in this WC is due to the fact that the contribution of the third generation quark inside the loop is dominant, as its degree of compositeness is the largest.

It is interesting to notice that the dependence on $\epsilon_f$ of the coefficients of dipole operators of Fig.~\ref{fig-Y3dip} is opposite to the one of Right-handed couplings of spin 1 particles, Fig.~\ref{fig-gRRud}. 
The behaviour of the Right-handed couplings $g^f_{R,12}$ is determined by their quadratic dependence on $\epsilon_f$, arising from the two insertions of elementary-composite mixing. The dependence of the dipole coefficients can be understood by considering a single generation: as the mass involves the combination $m\sim\epsilon_q Y \epsilon_f$, the dipole $d=\epsilon_1 Y^3 \epsilon_f$ scales with $\epsilon_f$ as $d\sim m^3\epsilon_q^{-2}\epsilon_f^{-2}$. Considering the presence of three generations adds the rotations into physical states, which are however mostly dominated by their diagonal part.

\subsection{Constraints from LHC}\label{sec-other}
Dijets at LHC give bounds on the compositeness scale of the light quarks~\cite{1212.6660,1312.3524,1410.8857,1703.09127,1703.09986,1806.00843,1911.03947}. Following Ref.~\cite{1706.03068}, see also~\cite{1706.03070}, the most stringent constraints arise form the 13 TeV run, that gives bounds on the Wilson coefficients of 4-fermion operators with light quarks, roughly of order: $(g^f_{L/R,11}v/m_*)^2\lesssim 10^{-3}$. For the present model one gets: $\lambda_C^{2 n_{f1}}\lesssim f/(8\ {\rm TeV})$, that if $f=1$~TeV is saturated for $n_{f1}\simeq 0.7$. Thus one has to demand $n_{f1}\gtrsim 0.7$, discarding the region of the solutions with large degree of compositeness of the Right-handed quarks of the first generation.

The particles associated with the radial part of the field $\Phi$, being electrically neutral and CP-even, can mix with the Higgs. As this field is a singlet, it has no effect on the oblique electroweak parameters, but it renormalizes the Higgs couplings to the SM fields. From Refs. \cite{ATLAS:2016neq,Dawson:2017vgm}, a combination of ATLAS and CMS results for single Higgs production gives a bound on the mixing angle: $\cos{\theta}>0.94$ at 95\% confidence level. Considering a renormalisable potential with a quartic coupling $\lambda_{H\Phi} |\Phi|^2 |H|^2$, in terms of the physical mass of the heavy state this bound leads to $M_\phi/\lambda_{H\Phi}\gtrsim 1.5$~TeV, where for simplicity we have assumed $v_\Phi\sim M_\phi\gg M_h$.

\section{The axion of composite Froggatt-Nielsen}\label{sec-axion}
There has been interest in composite axions in the last years~\cite{Alanne:2018fns,1911.09385,2007.10875,2012.09728}, see also Refs.~\cite{Ema:2016ops,Calibbi:2016hwq} for previous studies on the Froggatt-Nielsen axion. In the present paper the axion is associated with the NGB of the complex SM singlet: 
\begin{equation}
\phi=f_a e^{ia/f_a}+\dots
\end{equation}
with decay constant $f_a=\langle\phi\rangle$.
Usually $f_a$ is expected to be of order $m_*/g_*$, although perhaps it could be possible to differentiate these scales, if for example they arise from different sectors or are generated at different energies.

In the absence of explicit breaking of U(1)$_F$ in the elementary sector the axion remains massless, up to effects from anomalous breaking generated by QCD. In the presence of a composite sector with global SU(3)$_c$ symmetry, gauged by QCD, there can be large contributions to the axion mass that dominate over the IR ones~\cite{Gherghetta:2020keg}. Besides, there can also be contributions from the elementary sector. These effects make the composite axion rather dependent on the details of the model. In the following we will only focus on the properties of the axion that are independent of those details, and leave a deeper analysis of its phenomenology for the future.

Since the SM fermions are charged under U(1)$_F$, the axion is coupled to them, leading to a generalised DFSZ-type of model~\cite{Zhitnitsky:1980tq,Dine:1981rt}. 
By redefinition of the quarks: $f\to e^{i\gamma^5P_Fa/f_a}f$, the axion is removed from the Yukawa terms and axion-fermion interactions are generated in the kinetic term. In the mass basis we get:
\begin{equation}
{\cal L}\supset \frac{\partial_\mu a}{2f_a}\bar f_j\gamma^\mu(c^{Vf}_{jk}+c^{Af}_{jk}\gamma^5)f_k \ , \qquad f=u,d.
\end{equation}
The vector and axial-vector couplings are:
\begin{equation}
c^{V,Af}_{jk}=(\pm U^\dagger_{f_L}P_FU_{f_L}+U^\dagger_{f_R}P_FU_{f_R})_{jk} \ .
\label{cVA-eq}
\end{equation}
For non-universal $F$-charges $c^{V,Af}$ are non-diagonal and induce flavor transitions. For small mixing angles, to leading order these couplings can be approximated by:
\begin{equation}
(U^\dagger_{f}P_FU_{f})_{jk}=p_{fj}\delta_{jk}+(p_{fj}-p_{fk})\theta_{f,jk}+(p_{fj}+p_{fk}-2p_{fl})\theta_{f,jl}\theta_{f,kl} \ , \qquad l\neq j,k\ .
\end{equation}
By making use of Eqs.~(\ref{eq-thetaLjk}) and~(\ref{eq-thetaRjk}) in~(\ref{cVA-eq}) it is straightforward to obtain the size of $C^{V,Af}$ in the model.

The flavor violating effects of the axion depend on the decay constant $f_a$ and on the axion mass, whose values we did not need to fix for the analysis of the previous sections. Since these quantities can take a wide range of values, particularly the mass can vary over many orders of magnitude depending on features that are not generic, we postpone their analysis for a future work.

\section{Conclusions and discussions}\label{sec-conclusions}
We have built a model that realises flavor in the quark sector by the interplay of two paradigms: partial compositeness of the fermions and the Froggatt-Nielsen mechanism, by including the FN field in the composite sector, as well as the Higgs field. FN gives a well defined pattern of Yukawa couplings in the composite sector, determined by the charges of the composite operators under the global U(1)$_F$  symmetry. Partial compositeness is, as usual, realised by linear mixing of the elementary quarks with the composite sector, with the mixing controlling the interactions between both sectors. The two ingredients give rise to a rich pattern of flavor structures, that go beyond the usual scenario of APC. We have shown the basic rules for the determination of mixings and FN charges that lead to the masses and mixing angles of the quark sector of the SM.

We have chosen a set of solutions that lead to small mixing angles of Right-handed light quarks, and we have analysed the predictions for the flavor and CP violating processes that have the most stringent constraints. We have compared them with APC that, for a scale of composite states of order few TeV, is known to pass many flavor tests, but gives too large contributions to $\Delta F=2$ operators in the down sector and to diagonal dipole operators of light quarks, between others. We have found that Wilson coefficients of Left-handed $\Delta F=2$ operators are of the same size as APC, as expected for Left angles of CKM size, but for operators involving Right-handed light quarks we have found solutions that are suppressed compared with APC. Particularly interesting is the case of $Q_4^{sd}$, that is one of the most constrained operators, for which we have found solutions with a Wilson coefficient suppressed by $\lambda_C^2$. Another interesting case are the flavor diagonal dipole operators, also with solutions suppressed by $\lambda_C^2$ compared with APC.

For the solution considered we explored a range of charges and degrees of compositeness of the Right-handed quarks of the first and second generations. Flavor constraints select a preferred set of charges and mixings, with the second generation of Right-handed quarks uncharged, whereas the first generation charges of Left- and Right-handed chiralities are of ${\cal O}(1)$, and there is a window for the degree of compositeness of the Right-handed quarks of the first generation: $\epsilon_{d1}\simeq \lambda_C^{4-5}$ and $\epsilon_{u1}\simeq \lambda_C^{2-5}$. The compositeness of the down quark is determined by the bounds from $\Delta F=2$ processes in the $K$-system that prefer a larger $\epsilon_{d1}$, and $\Delta F =0$ dipole operators that prefer a smaller $\epsilon_{d1}$. A similar situation holds for the compositeness of the up quark, with constraints from $D$-system preferring a larger $\epsilon_{u1}$ and from $uuV$ dipole preferring a smaller $\epsilon_{u1}$. For the selected window of compositeness, dipole operators are suppressed at least by $\lambda_C^2$ and $C_4$ is suppressed at least by $\lambda_C$, relaxing $m_*\gtrsim 2.5-7.5$~TeV and $f$ below the TeV. There are also regions where $C_4$ is suppressed by $\lambda_C^{3/2}\sim 10^{-1}$. The $ccV$ dipole can only be suppressed if this quark is charged, being in tension with $\Delta F=2$ constraints that prefer zero charge. Concerning flavor violating $Z$ couplings, for a generic composite sector bounds from $B_s\to \mu\mu$ introduce tuning of order $\lambda_C^2$, that can be relaxed by the use of discrete symmetries protecting Left-handed down-type couplings.

Several questions were left open. At the level of bounds from flavor, we have not been able to find solutions that could lead to Left-handed mixing angles smaller than the CKM, either in the up- or in the down-sector, or a partial combination of them, while simultaneously passing bounds from $\Delta F=2$ processes. We have neither been able to find solutions that could suppress the mixing in the $B_s$ system, compared with APC. 
At a more theoretical level, it would be interesting to find a rationale for the values of the U(1)$_F$ charges that, although being of order one, are arbitrary and have been chosen to be multiples of the charge of the FN scalar field. We have not explored the flavor of the leptons in the proposed scenario. It would also be interesting to build a more predictive model of the composite sector, as a realisation in five dimensions. 
Last, we have not explored the different possibilities for the axion-like state of the model, that eventually could solve the strong CP problem and pass the axion flavor constraints.

\section*{Acknowledgements}
We thank Oscar Cat\`a for conversations that triggered this work. This project has been partially supported by CONICET Argentina with PIP-0299 and FONCyT with PICT-2018-03682.

\bibliographystyle{JHEP}
\bibliography{biblio_v2}

\end{document}